\documentclass[%
 reprint,
superscriptaddress,
 amsmath,amssymb,
 aps,
prl,
]{revtex4-1}

\usepackage[caption=false]{subfig}
\usepackage{graphicx}% Include figure files
\usepackage{dcolumn}% Align table columns on decimal point
\usepackage{bm}% bold math
\usepackage{color}   %May be necessary if you want to color links
\usepackage{hyperref}
\hypersetup{
	colorlinks=true, %set true if you want colored links
	linktoc=all,     %set to all if you want both sections and subsections linked
	linkcolor=blue,  %choose some color if you want links to stand out
	citecolor=blue,
	urlcolor=blue
}

\begin{document}

%\preprint{APS/123-QED}

\title{Thickness dependent electronic structure in WTe$_2$ thin films}% Force line breaks with \\
%\thanks{A footnote to the article title}%

\author{Fei-Xiang Xiang}
\email{feixiang.xiang@unsw.edu.au}
\affiliation{
	Institute for Superconducting and Electronic Materials, Australian Institute for Innovative Materials, University of Wollongong,
	Innovation Campus, North Wollongong, New South Wales 2500, Australia%\\ %This line break forced% with \\
}%
\affiliation{
 School of Physics, The University of New South Wales, Sydney NSW 2052, Australia
}%
% \altaffiliation[Also at ]{Physics Department, XYZ University.}%Lines break automatically or can be forced with \\
\author{Ashwin Srinivasan}%
\affiliation{
	School of Physics, The University of New South Wales, Sydney NSW 2052, Australia
}%

\author{Oleh Klochan}
\affiliation{
	School of Physics, The University of New South Wales, Sydney NSW 2052, Australia
}%

\author{Shi-Xue Dou}
\affiliation{
	Institute for Superconducting and Electronic Materials, Australian Institute for Innovative Materials, University of Wollongong,
	Innovation Campus, North Wollongong, New South Wales 2500, Australia%\\ %This line break forced% with \\
}%

\author{Alex R. Hamilton}
 \email{alex.hamilton@unsw.edu.au}
\affiliation{
	School of Physics, The University of New South Wales, Sydney NSW 2052, Australia
}%

\author{Xiao-Lin Wang}
 \email{xiaolin@uow.edu.au}
\affiliation{
	Institute for Superconducting and Electronic Materials, Australian Institute for Innovative Materials, University of Wollongong,
	Innovation Campus, North Wollongong, New South Wales 2500, Australia%\\ %This line break forced% with \\
}%

%\collaboration{CLEO Collaboration}%\noaffiliation

\date{\today}% It is always \today, today,
             %  but any date may be explicitly specified

\begin{abstract}
We study the electronic structure of WTe$_2$ thin film flakes with different thickness down to 11 nm. Angle-dependent quantum oscillations reveal a crossover from a three-dimensional (3D) to a two-dimensional (2D) electronic system when the sample thickness is reduced below 26 nm. The quantum oscillations further show that the Fermi pockets get smaller as the samples are made thinner, indicating that the overlap between conduction and valence bands is getting smaller and implying the spatial confinement could lift the overlap in even thinner samples. In addition, the quadratic magnetoresistance (MR) also shows a crossover from 3D to 2D behavior as the samples are made thinner, while gating is shown to affect both the quadratic MR and the quantum oscillations of a thin sample by tuning its carrier density.

%\begin{description}
%\item[Usage]
%Secondary publications and information retrieval purposes.
%\item[PACS numbers]
%May be entered using the \verb+\pacs{#1}+ command.
%\item[Structure]
%You may use the \texttt{description} environment to structure your abstract;
%use the optional argument of the \verb+\item+ command to give the category of each item. 
%\end{description}
\end{abstract}

%\pacs{Valid PACS appear here}% PACS, the Physics and Astronomy
                             % Classification Scheme.
%\keywords{Suggested keywords}%Use showkeys class option if keyword
                              %display desired
\maketitle

%\tableofcontents

Tungsten ditelluride, one of the transition metal dichalcogenides, has attracted a lot of attention recently. Bulk WTe$_2$ exhibits an unsaturated and extremely large magnetoresistance (MR) which could be used to design devices such as magnetic sensors \cite{nature13763}. It is believed that the magnetoresistance occurs due to the nearly equal electron and hole concentrations   \cite{nature13763,PhysRevLett.113.216601,PhysRevLett.114.176601,PhysRevLett.115.057202,EPL112.37009,PhysRevB.92.041104,0295-5075-110-3-37004,ncomms10847}, and forbidden backscattering due to strong spin-orbit coupling also plays an important role \cite{PhysRevLett.115.166601}. Surprisingly, although WTe$_2$ is a layered material, the anisotropy of the effective mass is as low as 2 at room temperature \cite{PhysRevLett.115.046602}. Quantum oscillations can be observed along three different crystal axes \cite{PhysRevLett.114.176601,PhysRevB.92.125152}, which makes it close to a three-dimensional (3D) electronical system instead of two-dimensional (2D) electronic system \cite{Viewpoint}. Bulk WTe$_2$ is also predicted to be a type-II Weyl semimetal, in which Lorenz invariance is absent and the Weyl point appears at the boundary of electron and hole pockets \cite{nature15768}. This prediction has triggered renewed interests in this material \cite{PhysRevB.94.121113,PhysRevB.94.241119,PhysRevB.94.195134,PhysRevB.94.161401,PhysRevB.94.121112,ncomms13142}. Furthermore, monolayer WTe$_2$ is predicted to be a nontrivial semimetal where both topological metallic edge states and 2D metallic bulk state are present~\cite{Science3466215}. Applying a small tensile strain, however, could lift the overlap of the conduction and valence bands and turn it into a 2D topological insulator~\cite{Science3466215}, where only the topological edge state is metallic \cite{PhysRevLett.95.146802,*PhysRevLett.95.226801,PhysRevLett.96.106802,*Science3141757,*Science3185851}.

The studies of the unusual MR of bulk WTe$_2$ has been extended to thin film samples~\cite{ncomms9892,PhysRevLett.117.176601,nanoscale818703,arXiv:1606.05756,arXiv:1608.04801,arXiv:1702.05876}. Electron and hole compensation is found in samples down to thickness around 10 nm, but the MR of thin films is suppressed significantly due to surface scattering from amorphous surface oxides \cite{nanoscale818703,arXiv:1606.05756,srep10013,SMLL201601207,2053-1583-4-2-021008}. Meanwhile the carrier density also has been observed to decrease as the samples are made thinner ~\cite{nanoscale818703,arXiv:1606.05756}, which is unexpected if it is assumed that thin film samples are simply thinner cases of 3D bulk samples. Recently optical and transport measurements in thin film samples with thickness from 9 to 11.7 nm suggests a band gap could open in monolayer sample and this is supported by band structure calculations \cite{adma.201600100}. Another study reported that semi-metallic monolayer samples become insulating in the 2D bulk with a topological nontrivial metallic state on their edge when the temperature is below $\sim$ 100 K \cite{arXiv:1610.07924}.  Therefore, to explore the transition from the 3D bulk state to the predicted 2D topological insulator, it is necessary to study the electronic structure of WTe$_2$ thin film samples when the sample thickness is approaching the monolayer limit.

Quantum oscillations due to Landau quantization are a powerful method to study band structure and Fermi surface of materials \cite{Shoenberg,Science329821,nphys1861,PhysRevB.92.035123}. Here we use the quantum oscillations observed in magneto-transport to study thickness-dependent electronic structure of WTe$_2$ thin films. Angle-dependent quantum oscillations reveal a 3D-to-2D crossover when the sample thickness is below 26 nm. The fast Fourier transform of the quantum oscillations further shows that the area of the Fermi pockets decreases as the sample is made thinner, suggesting that the overlap between the conduction band minimum and valence band maximum of the semi-metallic WTe$_2$ thin film becomes smaller. This not only explains the decease of carrier density as the sample is made thinner as reported in Refs.~\cite{nanoscale818703,arXiv:1606.05756} but also suggests it is possible to open a band gap and realize the 2D topological insulator in even thin samples, as predicted by theory\cite{Science3466215}. In addition, the quadratic MR also shows a crossover from 3D to 2D behavior as the samples are made thinner, while gating is shown to affect both the quadratic MR and the quantum oscillations of a thin sample by tuning its carrier density.

\begin{figure*}[t]
	\centering
	\subfloat[\label{fig:WTe2devicefig1a} ]{}
	\subfloat[\label{fig:WTe2devicefig1b} ]{}
	\subfloat[\label{fig:WTe2devicefig1c} ]{}
	\subfloat[\label{fig:WTe2devicefig1d} ]{}
	\subfloat[\label{fig:WTe2devicefig1e} ]{}
	\subfloat[\label{fig:WTe2devicefig1f} ]{}
	\subfloat[\label{fig:WTe2devicefig1g} ]{}
	
	\includegraphics[width=0.9\textwidth]{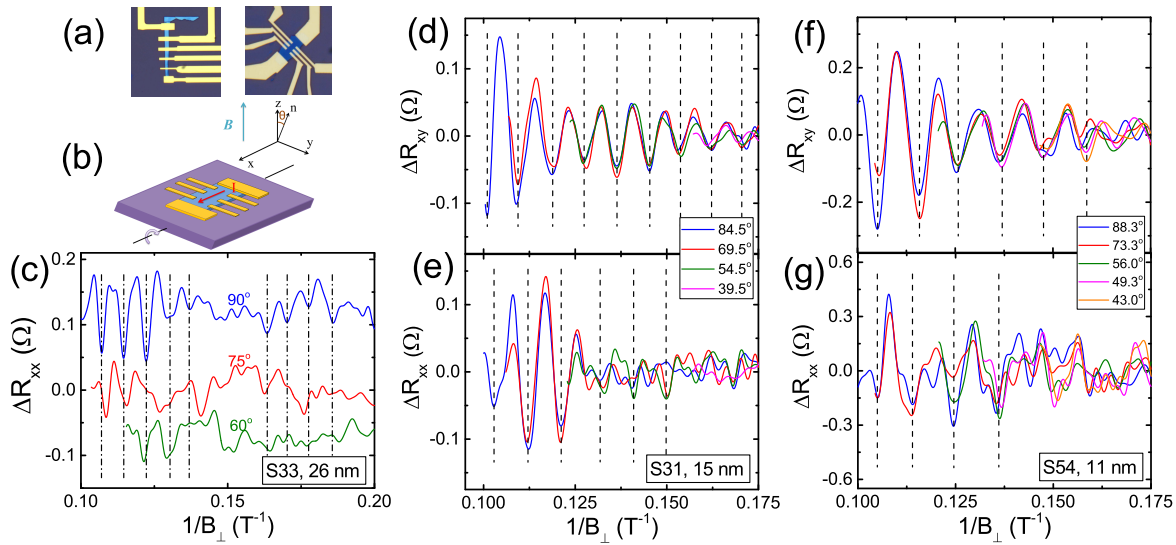}

	\caption{(a) Optical images of thick (right, S33) and thin (left, S54) samples. (b) Schematic diagram of angle-dependent measurement configuration. Magnetic field is parallel with $z$ direction, the rotation axis of substrate and current direction are parallel with $x$ axis, the $\theta$ is defined as an angle between $z$ axis and the normal, $n$, of sample surface. (c) Quantum oscillations $\Delta R_{xx}$ vs. $1/B_\perp$ for S33, the data traces at different angles are vertically offset for clarity. $B_\perp$ is perpendicular component of magnetic field, $B_\perp =B\sin \theta$. The black dash dot lines indicate the minima of $\Delta R_{xx}$ at $\theta=90$$^\circ$. (d), (e) and (f), (g) are the quantum oscillations $\Delta R_{xx}$ vs $1/B_\perp$ and $\Delta R_{xy}$ vs. $1/B_\perp$ for S31 and S54, respectively. The black dashed lines indicate the minima of $\Delta R_{xx}$ and $\Delta R_{xy}$. All the quantum oscillations $\Delta R_{xx}$ and $\Delta R_{xy}$ are obtained by subtracting polynomial background from $R_{xx}$ and $R_{xy}$ and are low pass filtered in $1/B$ scale to remove high frequency noise.}
	\label{fig:WTe2devicefig1}
\end{figure*}

WTe$_2$ thin films with different thickness are cleaved from a bulk WTe$_2$ single crystal by microexfoliation method onto a 285 nm SiO$_2$/Si substrate with alignment markers. E-beam lithography is used to fabricate the alignment makers, electrodes, and bonding pads. Before depositing Ti/Au electrodes with typical thickness 10nm/60nm, the contact areas are treated with Ar plasma to remove native oxides. To further reduce the contact resistance, the devices are annealed in a furnace with N$_2$ gas at 200 $^{\circ}$C for 2 hours. To reduce the oxidization of the sample surface, after cleaving, the thin films are only exposed to air before the contact area patterning with EBL and during the bonding of the devices to the chip carrier. At other times, the thin films are always covered by PMMA and stored in a vacuum desiccator or a N$_2$ glove box. Magneto-transport measurement was performed in an Oxford dilution fridge with an in-situ rotator \cite{RevSciInstr81113905} using standard low frequency lock-in techniques. Unless otherwise stated magneto-transport measurements were performed at $T=30$ mK. The sample thickness was measured by an atomic force microscope (AFM) after the magneto-transport measurement.

\begin{figure}[t]
	\centering
	
	\subfloat[\label{fig:WTe2devicefig2a} ]{}
	\subfloat[\label{fig:WTe2devicefig2b} ]{}
	\subfloat[\label{fig:WTe2devicefig2c} ]{}

	\includegraphics[width=0.5\textwidth]{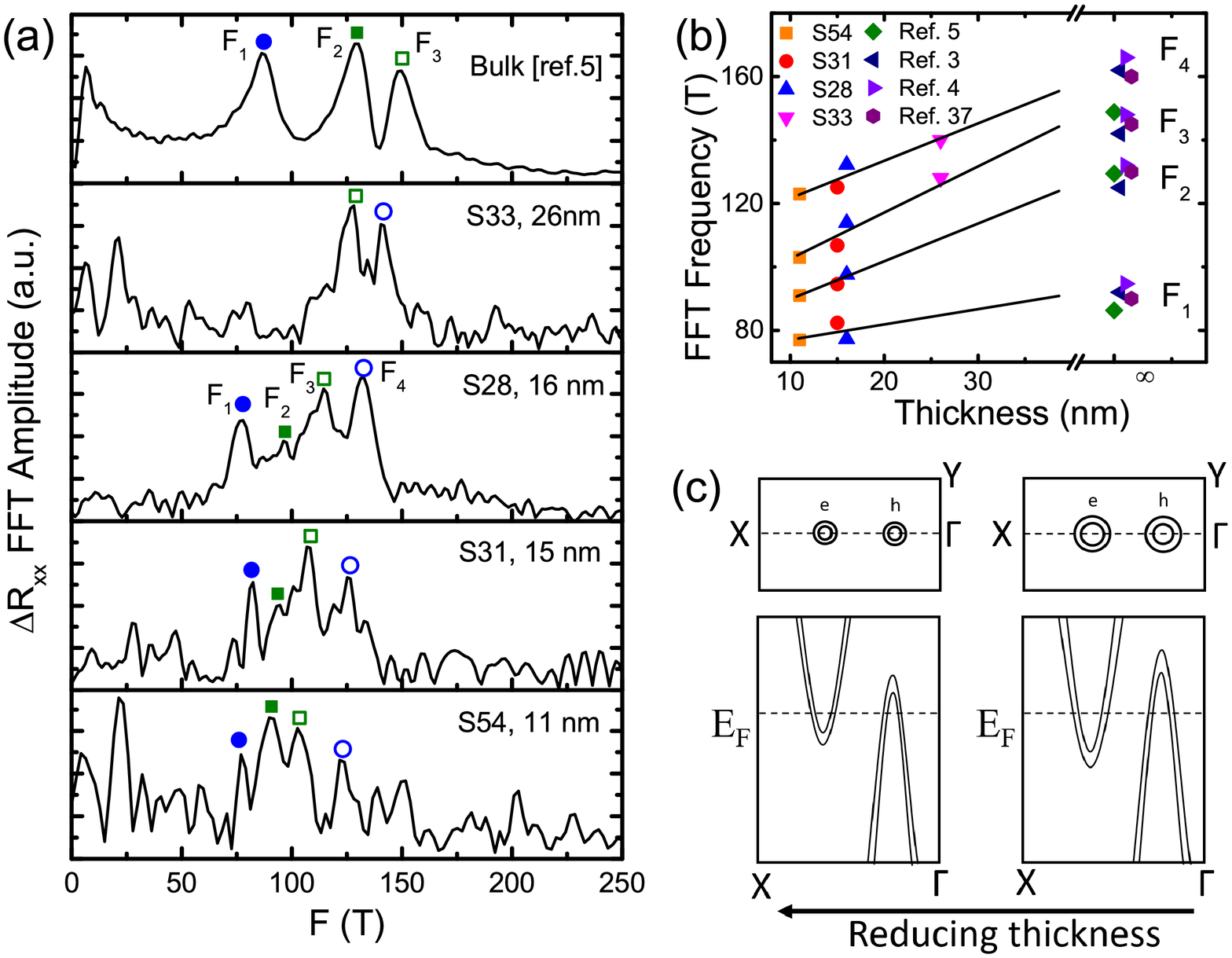}
	
	\caption{(a) Fast Fourier transform (FFT) spectrum of quantum oscillations, $\Delta R_{xx}$, of WTe$_2$ thin film samples with different thickness and bulk reference sample \cite{EPL112.37009}. The nominal $\theta = 90^\circ$. (b) The FFT frequency vs. sample thickness, the thickness of bulk from Refs.~\cite{PhysRevLett.114.176601,PhysRevLett.115.057202,EPL112.37009,PhysRevB.92.125152} is set as infinite compared with thin film samples. The solid lines are the guide lines which show the shift of frequency as thickness decreases. (c) Schematic diagrams showing the change of Fermi surfaces in half of Brillouin zone (top panels) and the band structure near Fermi surface along $\Gamma-X$ direction (bottom panels) as thickness is reduced (from right to left). }
	\label{fig:WTe2devicefig2}
\end{figure}

First, we show the WTe$_2$ thin film below 26 nm is a 2D system, contrary to previous suggestions that it is a 3D bulk system \cite{nanoscale818703,PhysRevLett.117.176601}. Figure~\ref{fig:WTe2devicefig1a} shows optical images of two typical devices, S33 (green) and S54 (blue). Figure~\ref{fig:WTe2devicefig1b} shows a schematic diagram of the measurement configuration for angle-dependent magneto-transport measurement. The current is driven through the $ab$ plane of the WTe$_2$ crystal, and when the magnetic field is rotated from out-of-plane to in-plane, it is always perpendicular to the current. We observe quantum oscillations related to the Landau quantization both in $R_{xx}$ and $R_{xy}$. Clear quantum oscillation signals $\Delta R_{xx}$ and $\Delta R_{xy}$ are obtained by subtracting a polynomial background from $R_{xx}$ and $R_{xy}$ as shown in Figs.~\ref{fig:WTe2devicefig1c}$-$\ref{fig:WTe2devicefig1g}. 

In bulk WTe$_2$ the Fermi surface is three dimensional, so quantum oscillations are observed for any magnetic field orientation, and have been studied for all three field orientations with respect to the crystal axes \cite{PhysRevLett.114.176601,PhysRevB.92.125152}. Only a small mass anisotropy is found for different transport directions \cite{PhysRevLett.115.046602}. In our thickest sample S33 (26 nm), we observe bulk behavior of the quantum oscillations when the sample is tilted in the applied magnetic field as shown in Fig.~\ref{fig:WTe2devicefig1c}. With the magnetic field perpendicular to the sample ($\theta = 90 ^\circ$, blue trace), there are clear oscillations in $\Delta R_{xx}$, periodic in $1/B_\perp$ which is the perpendicular component of magnetic field, $B_\perp =B\sin \theta$. However, the period and amplitude of these oscillations vary with the tilt angle $\theta$, indicating that the Fermi surface is three dimensional. In thinner samples, 15 nm or less, we find that both the period and amplitude of the quantum oscillations depend only on $B_\perp$, as shown in Figs.~\ref{fig:WTe2devicefig1d}$-$\ref{fig:WTe2devicefig1g}. This indicates that there is a crossover from a 3D to a 2D Fermi surface when the sample thickness is reduced from 26 nm (S33) to 15~nm~(S31).

Next we show another aspect of the changing electronic structure as the WTe$_2$ films are made thinner. Figure~\ref{fig:WTe2devicefig2a} shows a fast Fourier transform (FFT) spectrum of quantum oscillations in $\Delta R_{xx}$ from several samples, ranging from bulk to 11 nm thick. The FFT of the bulk sample obtained from our previous work at 2.5 K  shows three frequency peaks, 86.3 T, 130.2 T and 140.1 T \cite{EPL112.37009}. The absence of a fourth peak around 162 T reported in \cite{PhysRevLett.114.176601,PhysRevLett.115.057202,PhysRevB.92.125152} may be due to the relatively high measurement temperature and low mobility of carrier in that band. In order to resolve quantum oscillations from low mobility bands, all thin film samples were measured at millikelvin temperature and magnetic fields up to 10 T. We note that the quantum oscillation signals from the WTe$_2$ thin films are relative weak, reducing the signal to noise ratio in the FFT spectrum and hindering observation of the first two frequency peaks in S33. But for most samples (S28, S31 and S54) four frequency peaks can be observed. Sample S28 has the highest signal to noise ratio, so four strong frequency peaks can be clearly observed, as indicated by the squares and circles in Fig.~\ref{fig:WTe2devicefig2a}, which is consistent with the four frequencies observed in bulk samples \cite{PhysRevLett.114.176601,PhysRevLett.115.057202,PhysRevB.92.125152}. However, the peaks of the thin film samples shift to lower frequency compared to those of bulk samples.

To better illustrate the shift of frequency as the thickness is reduced, we plot the FFT frequency vs thickness as shown in Fig.~\ref{fig:WTe2devicefig2b}. The frequencies of bulk samples are from Refs.~\cite{PhysRevLett.114.176601,PhysRevLett.115.057202,EPL112.37009,PhysRevB.92.125152} and their thicknesses are set as infinite compared with the thin film samples. The solid lines are guides to the eye showing how each frequency shifts as the sample thickness is reduced. The frequencies $F_1$, $F_2$, $F_3$ and $F_4$  correspond to the frequency peaks indicated by the solid circle, solid square, open square and open circle in the Fig.~\ref{fig:WTe2devicefig2a}, respectively. It can be seen that there is a trend that the FFT frequency becomes smaller as the sample thickness is reduced. And recent work on a three layer WTe$_2$ sample shows even smaller quantum oscillation frequency \cite{PhysRevB.95.041410}. The frequency of quantum oscillations is related to the extremal cross-sectional areas of the Fermi surface perpendicular to the magnetic field. The lower frequencies correspond to smaller area of Fermi pockets. One possibility is that the reduction in the size of the Fermi pockets is caused by a change in the Fermi level. However, since both the electron and hole pockets are simultaneously occupied in WTe$_2$, changes in the Fermi level would cause the area of some Fermi pockets to increase while others decrease as the samples are made thinner. This is not the case here. The other possibility is change of band structure as illustrated in the Fig.~\ref{fig:WTe2devicefig2c}, which shows schematic diagrams of how a change of band structure could decrease the size of both electron and hole pockets. As the thickness of the samples is reduced (right to left), the conduction band minima and valence band maxima move apart, (bottom panel of Fig.~\ref{fig:WTe2devicefig2c}), which reduces the size of both electron and hole pockets as illustrated in the top panel. A similar confinement-induced band structure change has been observed in MoS$_2$, where the indirect band gap is changed from a bulk value of 1.29 eV to over 1.90 eV \cite{PhysRevLett.105.136805}, and in MoTe$_2$, a relative compound of WTe$_2$, where few layer samples shows band gap opening \cite{nphys3314}.This indicates it may be possible to observe band gap opening in even thinner WTe$_2$ thin films, which is consistent with two recent works on the band gap opening in monolayer samples \cite{adma.201600100,arXiv:1610.07924}.

Because the carrier density is proportional to the size of the Fermi pocket, a decrease of the Fermi pocket size should also result in a decrease of carrier density. We calculate the 2D carrier density from quantum oscillation frequencies in our high signal to noise ratio sample S28 (16 nm) by $n=g_sg_veF/h$, is $4.03 \times 10^{13}$ cm$^{-2}$ where $g_s=2$ and $g_v=2$ are spin and valley degeneracy, $e$ is electron charge, $h$ is Plank constant and $F$ is the sum of $F_1 = 77$ T, $F_2 = 98$ T, $F_3 = 114$ T and $F_4 = 125$ T. This corresponds to a 3D carrier density $2.69 \times 10^{19}$ cm$^{-3}$, very close to $3 \times 10^{19}$ cm$^{-3}$ from the 13 nm sample of Ref.~\cite{arXiv:1606.05756}, and similar to $7.5 \times 10^{19}$ cm$^{-3}$ from the 15.6 nm sample of Ref.~\cite{nanoscale818703}. Therefore, the change of Fermi pocket size explains the unexpected reduction of carrier density as the samples become thinner as reported in Refs.~\cite{nanoscale818703,arXiv:1606.05756}.
 
\begin{figure}[t]
	\centering
	
	\subfloat[\label{fig:WTe2devicefig3a} ]{}
	\subfloat[\label{fig:WTe2devicefig3b} ]{}
	\subfloat[\label{fig:WTe2devicefig3c} ]{}
	\subfloat[\label{fig:WTe2devicefig3d} ]{}
	\subfloat[\label{fig:WTe2devicefig3e} ]{}
	\subfloat[\label{fig:WTe2devicefig3f} ]{}

	\includegraphics[width=0.5\textwidth]{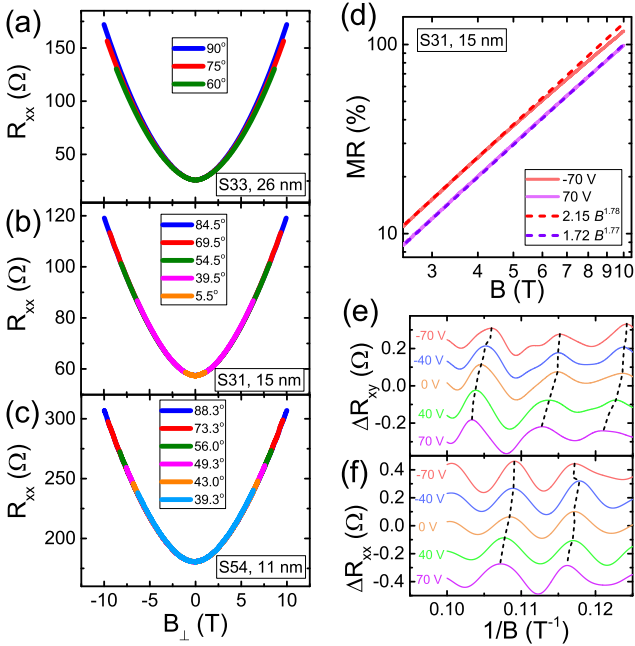}
	
	\caption{(a)$-$(c) $R_{xx}$ as a function of perpendicular component of magnetic field at different angles for sample S33, S31 and S54, respectively. The angles are the same as those in Fig.~\ref{fig:WTe2devicefig2}. (d) MR of S31 in log-log scale at $V_{bg}=$ 70 V and -70 V and nominal $\theta = 90^\circ$. The solid lines are experiment data and the dashed lines are power law fittings. (e),(f) Quantum oscillations from $R_{xy}$ and $R_{xx}$ of S31 at different $V_{bg}$. The oscillation traces are vertically offset for clarity. The dashed lines are the guides to eye which shows the shift of oscillation maxima.}
	
\end{figure}

Now we discuss the impact of thickness-dependent electronic structure of WTe$_2$ on its MR. In bulk samples, the angle-dependent $R_{xx}$ as a function of magnetic field exhibits a 3D-like scaling law, $R(B,\theta)=R(\epsilon_\theta B)$ with $\epsilon_\theta =(\cos^2\theta + \gamma^{-2}\sin^2\theta)^{1/2}$ \cite{PhysRevLett.115.046602}. In contrast, here we find $R_{xx}$ of the thin samples, S31 and S54, depends only on the perpendicular component of the magnetic field (see Figs.~\ref{fig:WTe2devicefig3b} and \ref{fig:WTe2devicefig3c}) while the thick sample S33 does not: $R_{xx}$ traces  at different angles deviates from each other at high magnetic field (see Fig.~\ref{fig:WTe2devicefig3a}). This indicates that the thin samples are 2D, while the thick sample are not, which is consistent with the angle-dependent quantum oscillation results in Fig.~\ref{fig:WTe2devicefig2}. 

The unsaturated MR in WTe$_2$ is believed to be due to perfect electron and hole compensation, as described by the two-band model \cite{nature13763,PhysRevLett.113.216601,PhysRevLett.114.176601,PhysRevLett.115.057202,EPL112.37009,PhysRevB.92.041104,0295-5075-110-3-37004}. If the electron and hole density are not equal, the MR will show a deviation at certain field, and the larger the difference between electron and hole density, the smaller field at which the deviation occurs \cite{nature13763,PhysRevB.95.041410}. The decrease of the carrier density as the sample thickness is reduced, as observed in Fig.~\ref{fig:WTe2devicefig2}, implies it is possible to use an electric field to tune the carrier density in a thin sample, which could provide a method to test the origin of the peculiar MR in WT$_2$. In the sample S31 (15 nm), we measure the MR, defined as $(R_{xx}(B)-R_{xx}(0))/R_{xx}(0)$, at different back-gate voltages ($V_{bg}$) from 70 V to -70 V. Figure~\ref{fig:WTe2devicefig3d} shows MR at 70 V and -70 V. While the MR at 70 V is well described by a simple power law up to 10 T, closed to the MR$=\mu_e\mu_h B^2$ obtain from two-band model, the MR at -70 V starts to deviate from the power law at 5 T. For $V_{bg}=0$ V this deviation occurs at 8 T (See Fig.~S4(c) in Supplementary Materials~\cite{Supplementary}). Because a positive (negative) $V_{bg}$ decreases (increases) the hole density and increases (decreases) the electron density, it means that the electron and hole density are nearly equal at 70 V. When $V_{bg}$ is changed from 70 V to ~-70 V, the Fermi level shifts down and the total electron density gradually become smaller than the total hole density. In addition, we observed that the back-gate also changes the quantum oscillations of S31 as shown in Figs.~\ref{fig:WTe2devicefig3e} and \ref{fig:WTe2devicefig3f}. It can be seen that the maxima of quantum oscillations as indicated by the dashed lines in $\Delta R_{xx}$ and $\Delta R_{xy}$ gradually shift to lower magnetic field as the $V_{bg}$ is changed from 70 V to -70 V. This indicates that the back-gate start to tune the carrier density, which consistent with the observation in Fig.~\ref{fig:WTe2devicefig3d}. With even thinner, the back-gate could be more effective to change the Fermi level, as observed in Refs.~\cite{PhysRevB.95.041410,acs.nanolett.6b04194}.

In summary, we observed thickness-dependent electronic structure in WTe$_2$ thin films, by studying quantum oscillations from magneto-transport measurements. One finding is a crossover from a 3D to a 2D electronic system when the sample thickness is less than 26 nm, contrary to the assumptions in previous studies \cite{PhysRevLett.117.176601,nanoscale818703}. FFT analysis shows that the size of the Fermi pockets becomes smaller as the samples are made thinner, indicating the overlap between conduction band minima and valence band maxima becomes smaller. This explains why the carrier density decreases as the sample thickness is reduced~\cite{nanoscale818703,arXiv:1606.05756}, and implies it is promising to open a band gap in even thinner samples and realize the 2D topological insulator \cite{Science3466215}. In addition, quadratic MR also shows a crossover from 3D to 2D behavior as the samples are made thinner, while gating is shown to affect both the quadratic MR and the quantum oscillations of a thin sample by tuning its carrier density.

\begin{acknowledgments}
We thank Shouyi Xie, Sven Rogge, Haifeng Feng and Yi Du for technical assistance, and thank Menno Veldhorst for helpful discussions. We acknowledge support from the ARC Discovery Project Scheme (DP160100077 and DP130102956) and an ARC Professional Future Fellowship (FT130100778). This work was performed in part using facilities of the NSW and ACT Nodes of the Australian National Fabrication Facility.
\end{acknowledgments}

\bibliography{WTe2references}% Produces the bibliography via BibTeX.

\end{document}

%\preprint{APS/123-QED}

\title{Supplementary Materials for "Change of electronic structure in WTe$_2$ thin film"}% Force line breaks with \\
%\thanks{A footnote to the article title}%

\author{Fei-Xiang Xiang}
%\email{f.xiang@unsw.edu.au}
\affiliation{
	Institute for Superconducting and Electronic Materials, Australian Institute for Innovative Materials, University of Wollongong,
	Innovation Campus, North Wollongong, New South Wales 2500, Australia%\\ %This line break forced% with \\
}%
\affiliation{
	School of Physics, The University of New South Wales, Sydney NSW 2052, Australia
}%
% \altaffiliation[Also at ]{Physics Department, XYZ University.}%Lines break automatically or can be forced with \\
\author{Ashwin Srinivasan}%
% \email{Second.Author@institution.edu}
\affiliation{
	School of Physics, The University of New South Wales, Sydney NSW 2052, Australia
}%

\author{Oleh Klochan}

\affiliation{
	School of Physics, The University of New South Wales, Sydney NSW 2052, Australia
}%

\author{Shi-Xue Dou}
\affiliation{
	Institute for Superconducting and Electronic Materials, Australian Institute for Innovative Materials, University of Wollongong,
	Innovation Campus, North Wollongong, New South Wales 2500, Australia%\\ %This line break forced% with \\
}%

\author{Alex R. Hamilton}
%\email{alex.hamilton@unsw.edu.au}
\affiliation{
	School of Physics, The University of New South Wales, Sydney NSW 2052, Australia
}%

\author{Xiao-Lin Wang}
%\email{xiaolin@uow.edu.au}
\affiliation{
	Institute for Superconducting and Electronic Materials, Australian Institute for Innovative Materials, University of Wollongong,
	Innovation Campus, North Wollongong, New South Wales 2500, Australia%\\ %This line break forced% with \\
}%

%\collaboration{CLEO Collaboration}%\noaffiliation

\date{\today}% It is always \today, today,
             %  but any date may be explicitly specified

%\pacs{Valid PACS appear here}% PACS, the Physics and Astronomy
                             % Classification Scheme.
%\keywords{Suggested keywords}%Use showkeys class option if keyword
                              %display desired
\maketitle

%\tableofcontents

\begin{figure*}[h]
	\centering

	\includegraphics[width=0.9\textwidth]{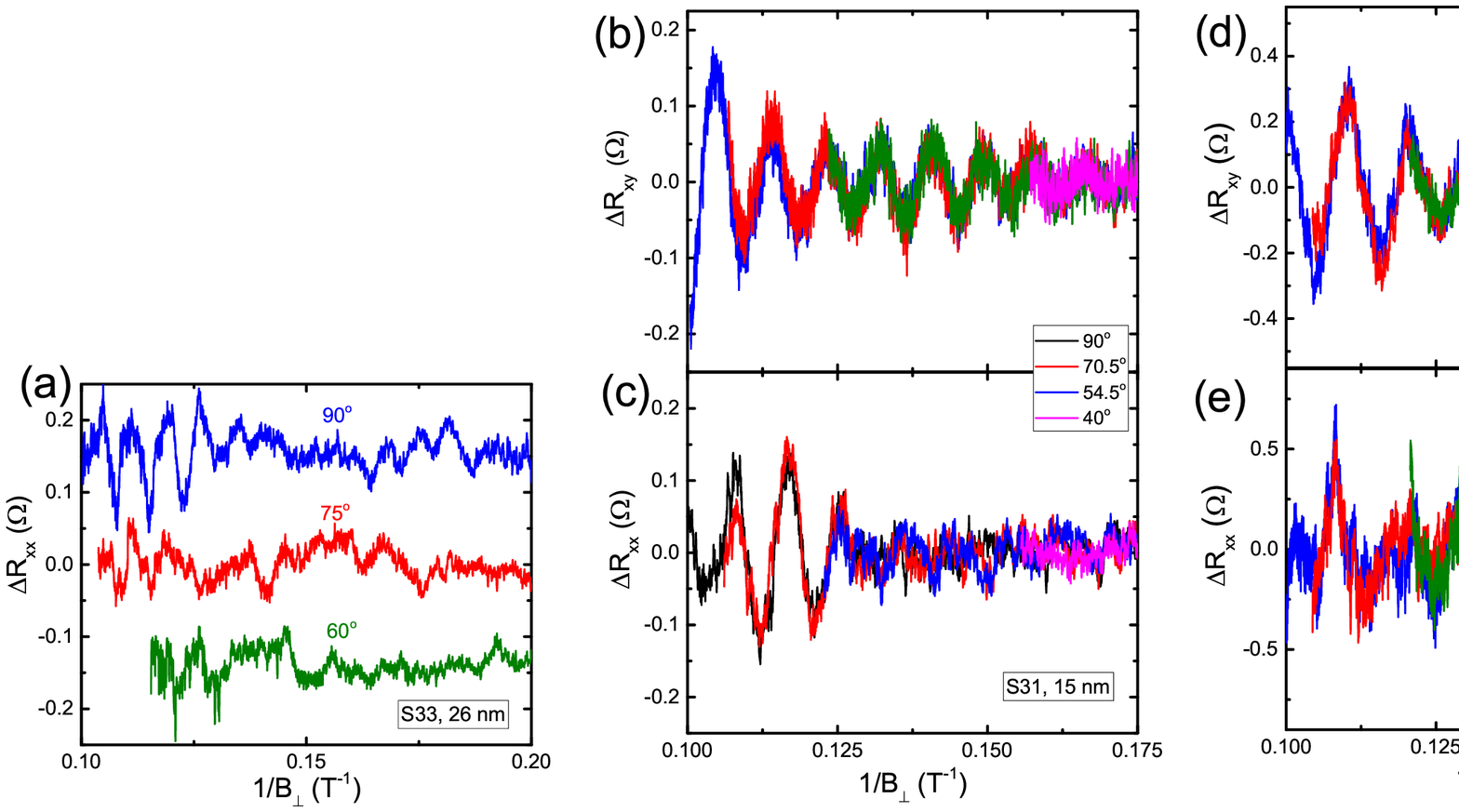}
	
		\caption{ Quantum oscillation signals from S33, S31 and S54 before low pass filtering. 
		}
	\label{fig:SM_angle_oscillation}
\end{figure*}

\begin{figure*}[h]
	\centering
	\subfloat[\label{fig:SM_MR_thickness_a} ]{}
	\subfloat[\label{fig:SM_MR_thickness_b} ]{}
	\subfloat[\label{fig:SM_MR_thickness_c} ]{}
	\subfloat[\label{fig:SM_MR_thickness_d} ]{}
	\subfloat[\label{fig:SM_MR_thickness_e} ]{}
	\subfloat[\label{fig:SM_MR_thickness_f} ]{}

	\includegraphics[width=0.9\textwidth]{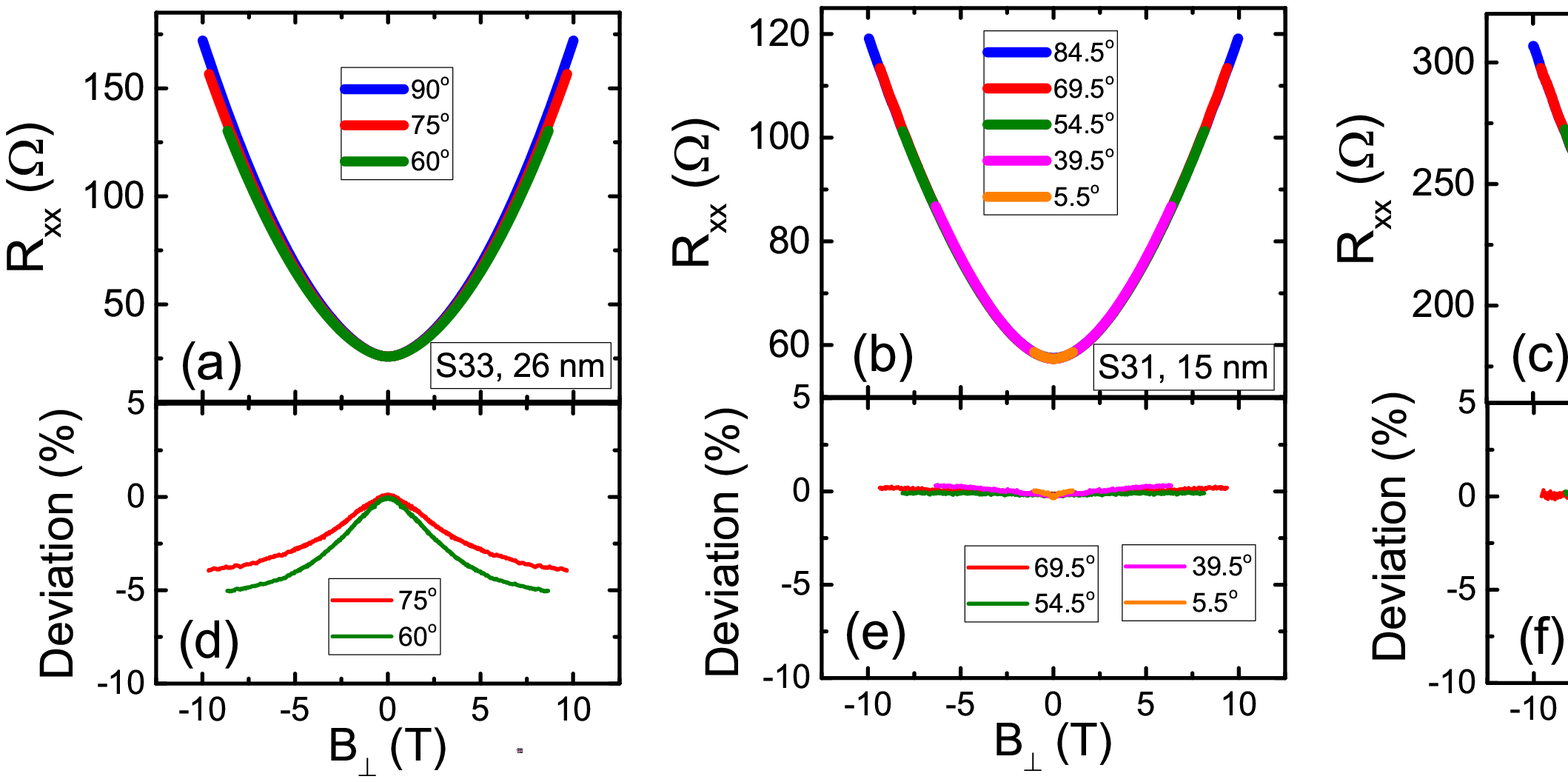}

	\caption{(a), (b) and (c) are perpendicular magnetic field dependent $R_{xx}$ at different angles for S33, S31 and S54.  (d), (e) and (f) are the deviation of MR as a function of perpendicular component of magnetic field from the largest angles, defined as $(R_{xx}(\theta,B)-R_{xx}(90^\circ,B))/R_{xx}(\theta_0,B)$ , where $\theta$ is the angle, $B$ is the magnetic field and $\theta_0$ is the largest angle for each measurement, $\theta_0 =90^\circ$, $84.5^\circ$ and $88.3^\circ$ for S33, S31 and S54, respectively.} 
	
	\label{fig:SM_MR_thickness}
\end{figure*}

\begin{figure*}[h]
	\centering

	\includegraphics[width=0.9\textwidth]{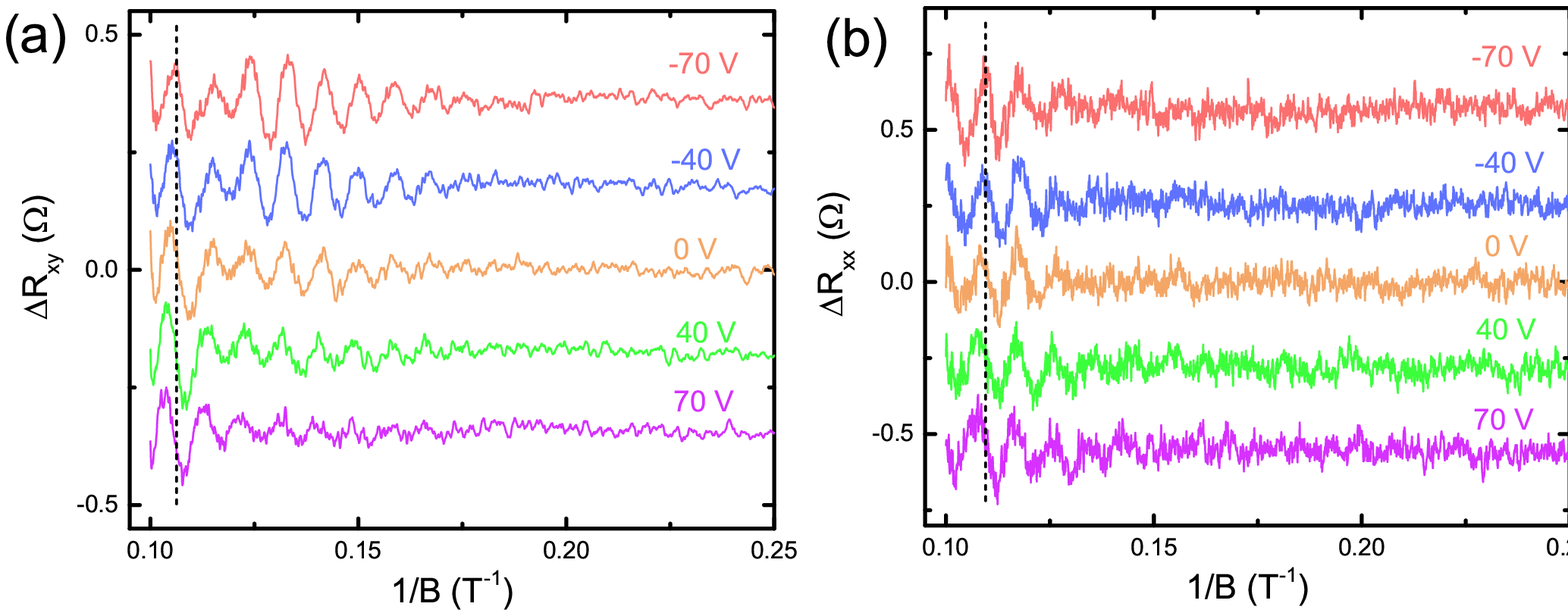}

	\caption{ (a), (b) are quantum oscillation signals $\Delta R_{xy}$ and $\Delta R_{xx}$ of S31 at different back-gate voltages before low pass filtering. The dash lines indicate one of the oscillation maxima in $\Delta R_{xy}$ and $\Delta R_{xx}$ at $V_{bg}=$ -70 V and along the lines the shift of oscillations can be observed. The deviation of the thin samples are almost flat line while the deviation of thick sample increase as magnetic field increase.
}
	\label{fig:SM_gate_dependence}
\end{figure*}

\begin{figure*}[h]
	\centering

	\includegraphics[width=0.9\textwidth]{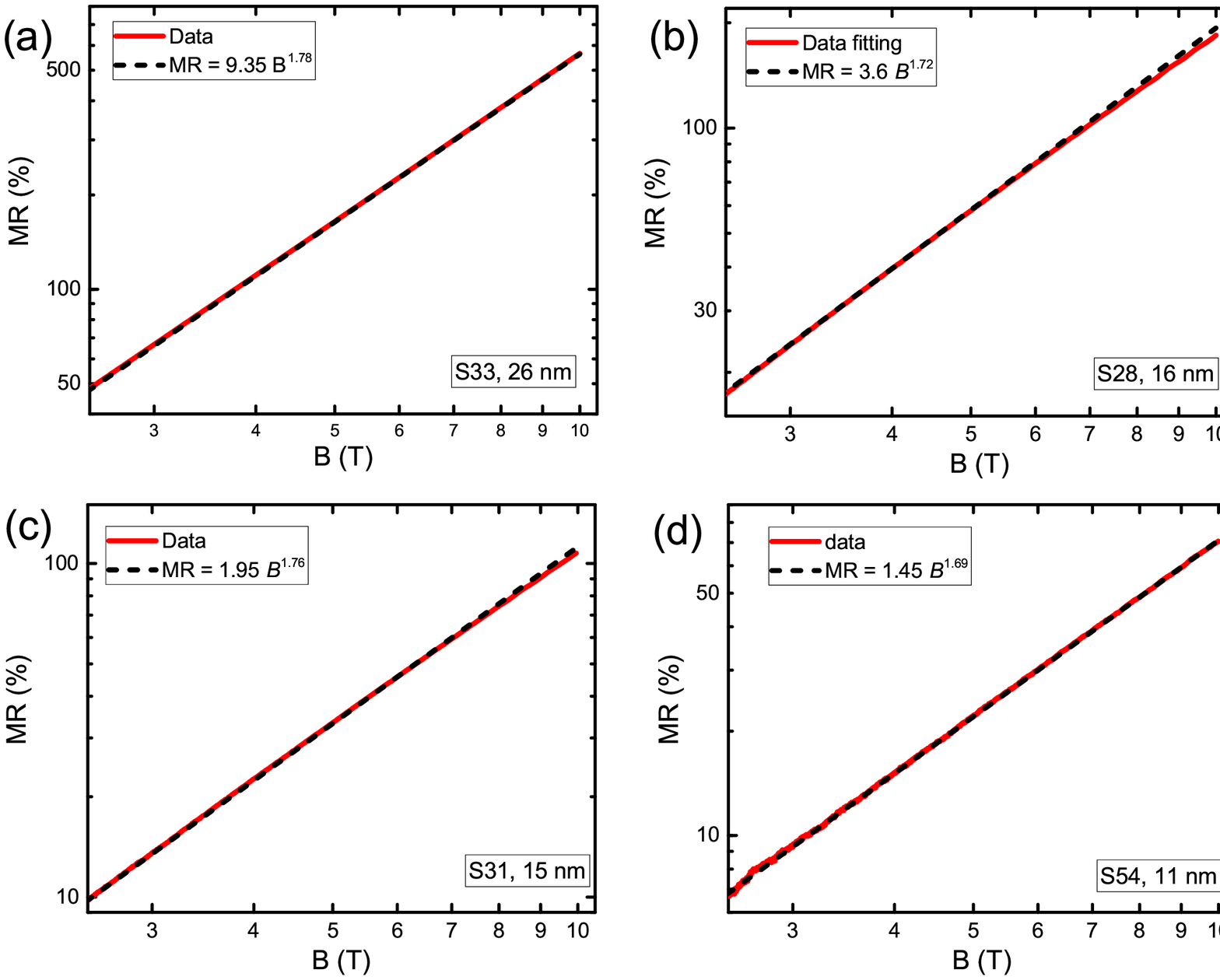}

	\caption{ (a), (b), (c) and (d) are the power law fitting of MR in log-log format from 2.5 to 10 T of sample S33, S28, S31 and S54, respectively. No back-gate voltage is applied. The red solid lines are experiment data and black dash lines are the fitting with power law.}
	\label{fig:SM_MR_fitting}
\end{figure*}

\begin{figure*}[h]
	\centering

	\includegraphics[width=0.6\textwidth]{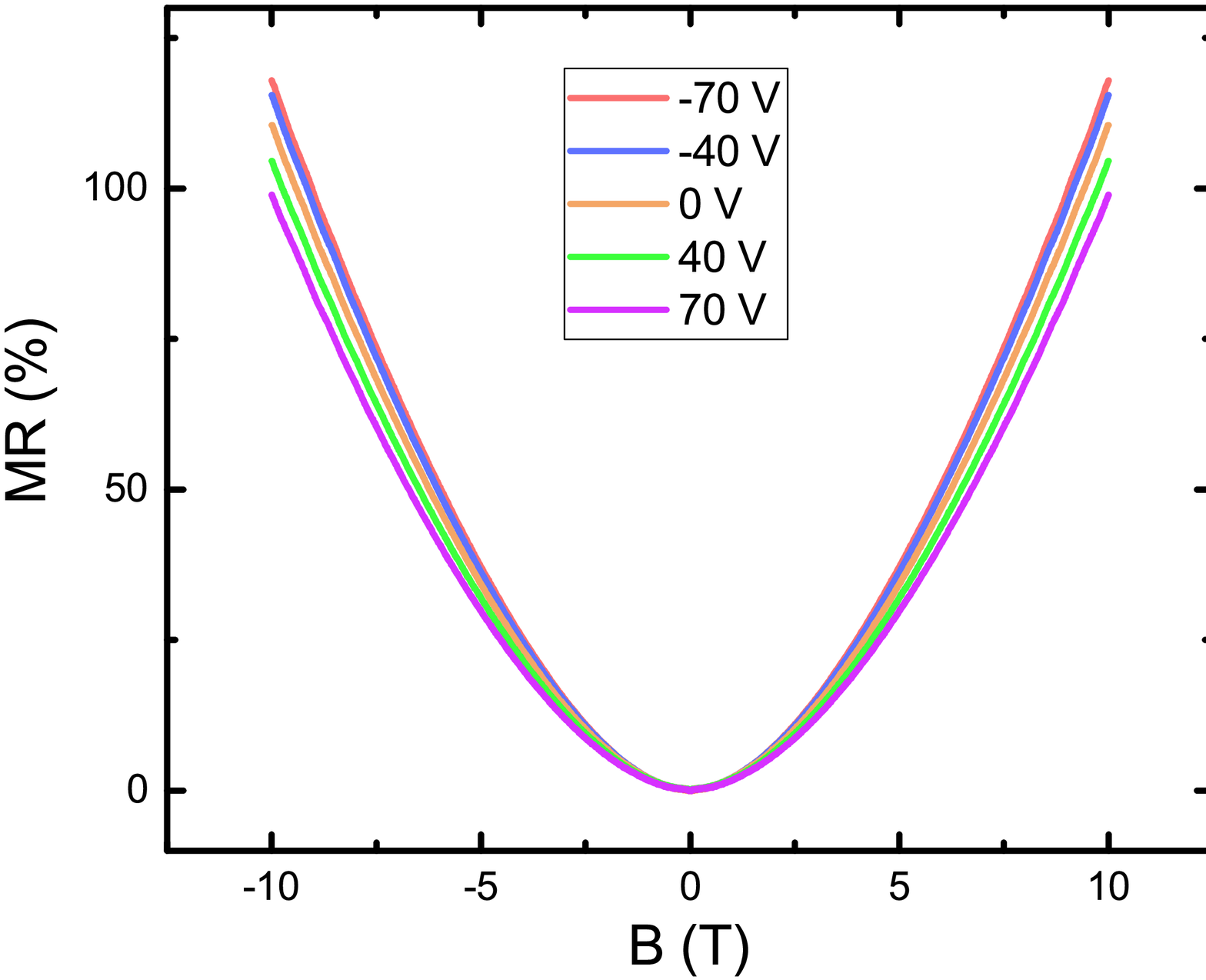}

	\caption{ MR of sample S31 in linear scale at different $V_{bg}$.}
	\label{fig:SM_MR_different_Vbg}
\end{figure*}

%\bibliography{references}% Produces the bibliography via BibTeX.